\documentclass[aps,prd]{revtex4}
\usepackage[colorlinks=true, pdfstartview=FitV, linkcolor=blue, citecolor=red, urlcolor=magenta]{hyperref}
\usepackage{graphicx}
\usepackage{latexsym}
\usepackage{amsmath}
\usepackage{amsfonts}
\usepackage{amssymb}
\usepackage{verbatim}

\newcommand{\be}{\begin{equation}}
\newcommand{\ee}{\end{equation}}
\newcommand{\bea}{\begin{eqnarray}}
\newcommand{\eea}{\end{eqnarray}}

\newcommand{\bb}{\bibitem}

\def\bb{\bibitem}

\def\bb{\bibitem}
\newcommand{\ben}{\begin{eqnarray}}
\newcommand{\een}{\end{eqnarray}}

\begin{document}

\title{A domain wall description of brane inflation and observational aspects}

\author{$^{1}$R. M. P. Neves}
\email{raissaeter@yahoo.com.br}
\author{$^{1}$F.F. Santos}
\email{fabiano.ffs23@gmail.com}
\author{ $^{1,2}$F. A. Brito}
\email{fabrito@df.ufcg.edu.br}

\affiliation{$^{1}$Departamento de F\'isica, Universidade Federal da Para\'iba, Caixa Postal 5008, 58051-970 Jo\~ao Pessoa, Para\'iba, Brazil}
\affiliation{$^{2}$Departamento de F\'{\i}sica, Universidade Federal de Campina Grande
Caixa Postal 10071, 58429-900 Campina Grande, Para\'{\i}ba, Brazil}

\begin{abstract}
We consider a brane cosmology scenario by taking an inflating 3D domain wall immersed in a five-dimensional Minkowski space in the presence of a stack of $N$ parallel domain walls. They are static BPS solutions of the bosonic sector of a 5D supergravity theory. However, one can move towards each other due to an attractive force in between driven by bulk particle collisions and {\it resonant tunneling effect}. The accelerating domain wall is a 3-brane that is assumed to be our inflating early Universe. We analyze this inflationary phase governed by the inflaton potential induced on the brane. We compute the slow-roll parameters and show that the spectral index and the tensor-to-scalar ratio are within the recent observational data.
\end{abstract}
\pacs{XX.XX, YY.YY} \maketitle


\section{Introduction}

Inflationary cosmological scenarios were proposed by Starobinsky, Sato, Guth \cite{Guth, sta,sato} and Linde \cite{Linde}. This phase of the Universe has been supported by Cosmic Microwave Background (CMB), discovered by Penzias and Wilson in 1964  \cite{Penzias} and verified accurately by COBE (Cosmic Background Explorer), WMAP (Wilkinson Microwave Anisotropy Probe) and PLANCK. The observations of the CMB have shown to develop enormous importance in modern cosmology concerning constrain several models that have emerged as an attempt to explain the expansion of the Universe \cite{Lindee, Costa, Santos,bbq-2007}]. Thus, the inflationary scenarios have been severely constrained by the recent data from the Planck collaboration \cite{Plan, Planck, Planckk}.
An interesting possibility is to associate the models that describe the acceleration of the Universe, such as during inflation and dark energy phases, to the scenario known as the Dvali-Tye brane inflation \cite{Dvali}. In this way, we consider our Universe as a 3-brane embedded in a 5D Minkowski spacetime \cite{Dvali, Brito} that undergoes an accelerated expansion due to the presence of an induced scalar potential for a scalar field that corresponds to inter-brane distance. This is the direction of looking from the inflationary scenario in the realm of fundamental theories such as string theories and effective limits, namely supergravity --- for a recent study on this issue see \cite{KL}. In the latter case, the bulk is asymptotically AdS$_5$ space-time. The effective inflaton potential in such theories is induced as the {\it radion} potential which comprises the inter brane potential as a function of the distance in between. The induced four-dimensional radion/inflaton appears naturally in AdS$_5$ bulk space due to modes that are integrated out \cite{RS,GW,MH}.
In the limit of Minkowski bulk space, however, it is also possible to find an inter brane potential in five-dimensional bulk that can induce a four-dimensional inflaton potential on the brane. In this scenario, one can take into account several forces among the branes \cite{Dvali}. In the present study, we investigate the realization of the Dvali-Tye scenario in the context of domain wall solutions in a 5D gravity coupled to scalar fields in a way that can be viewed as the bosonic sector of a particular supergravity theory in five dimensions. We shall consider a particular force due to elastic collisions of bulk particles with the branes. Besides, one can also consider other forces of gravitational and electromagnetic nature. We shall focus on the electromagnetic case to address issues in our setup. Due to the resonant tunneling effect placed between the branes, favoring the transmission rate in contrast to the reflection rate, it is expected to exist an attractive force in between whose magnitude increases (decreases) as the inter distance increases (decreases) --- Fig.~\ref{figs-N-branes}.  An analogous behavior in optical systems can be found in the optical spring phenomenon in a Fabry-Perot cavity \cite{corbitt}.

This paper is organized as follows. In Sec.~\ref{sec01}, we introduce supergravity inspired model from which we will find 3D domain wall solutions that represent the 3-branes. Sec.~\ref{inflaton}, we present a brane scenario in which we explore forces acting to the brane due to elastic collisions of bulk particles that can produce an acceleration in our Universe. We also discuss the presence of an electric field between parallel branes. After considering these forces, we deduce our potential induced on the brane. Sec.~\ref{sec02}, we consider such a model to determine the main parameters that govern the inflationary phase and discuss the results by making comparisons with the recent Planck data. Finally in Sec.~\ref{sec04}, we summarize the main results. 

\section{The model}\label{sec01}

We consider the scalar bosonic sector of {an effective} supergravity theory in 5D, which can be thought of as a theory that comes from another fundamental theory via compactification which Lagrangian is given by \cite{Brito,Cvetic}:
\bea\label{1}
	e^{-1}{\cal L}_{sugra}={-\frac{1}{4}M^3_*R_{(5)}}+G_{AB}\partial_m\phi^A\partial^m\phi_B-\frac{1}{4}G^{AB}\frac{\partial W(\phi)}{\partial\phi^A}\frac{\partial W(\phi)}{\partial\phi^B}+\frac{1}{3}\frac{1}{M^3_*}W(\phi)^2,
\eea 
{where $G_{AB}$ is the metric on the real scalar field space and $A,B=1,2,...$ label the number of scalar and fermion fields 
matching the number of supersymmetric degrees of freedom. } 
$e=|\det g_{mn}|^{1/2}$, $R_{(5)}$ is the Ricci scalar and $1/M_{*}$ represents the five-dimensional Planck length.

{The Lagrangian (\ref{1}) is the five-dimensional version of a more general class of effective supergravity Lagrangians in $D$ dimensions, which up to four-fermions terms is invariant under the following supersymmetric transformations  \cite{gibbons-lambert,yoon, lambert, nascimento} where only gravity, scalar fields $\phi_A$ and their superpartners $\psi_m$ and $\lambda_A$, respectively, are turned on
\ben 
\label{transf2-0}
\delta e_m^n&=& -\bar{\epsilon}\, \Gamma^n {\psi}_m+c.c.,\\ 
\label{transf1-0}
\delta\phi_A&=&\bar{\epsilon}\, \lambda_A+c.c.,\\
\label{transf1}
\delta\psi_m&=&\nabla_m\epsilon+\kappa^{D-2}W\Gamma_m\epsilon,\\
\label{transf2}
\delta\lambda_A&=&\left(-\frac12G_{AB}\Gamma^{m}\partial_m\phi^B+W_{3A}\right)\epsilon,
\een 
where, $m,n=0,1,...,D-1$, $\kappa=1/M_*$, $\epsilon$ is a local supersymmetry parameter, 
\ben
W_{3A} &=& (D-2)\frac{\partial W}{\partial\phi^A},
\een
and the scalar potential has the general form
\ben
V(\phi)&=&4(D-2)^2\left[G^{AB}\frac{\partial W(\phi)}{\partial\phi^A}\frac{\partial W(\phi)}{\partial\phi^B}-\left(\frac{D-1}{D-2}\right)\kappa^{D-2}W(\phi)^2\right].
\een
In the following analysis we will restrict the scalar field space to a flat two dimensional manifold, by freezing out the remaining degrees of freedom, keeping only the bosonic sector $\phi_A=(\phi_1,\phi_2)$ that corresponds to its fermionic superpartner $\lambda_A=(\lambda_1,\lambda_2)$. {We can associate to such a manifold, a {minimal `Kaehler-like potential'} $K=\phi_A\phi_A$},  such that $G_{AB}=2\delta_{AB}$. For later use, we anticipate that we can turn on further internal scalar degrees of freedom 
and recast their dynamics in terms of  a complex field.

\subsection{Domain wall solutions}

For our purpose of finding supersymmetric domain walls we assume without loss of generality the 5D spacetime metric as follows}
\be\label{5d-metric}
ds^2=e^{2A(x_5)}\eta_{\mu\nu}dx^\mu dx^\nu+dx_5^2, \qquad \mu,\nu=0,1,2,3.
\ee
For scalar fields depending only on the fifth dimension $x_5$ that represents the coordinate transverse to the brane and using the fact that the equations of motion of such systems {can be solved by solutions of first-order equations obtained from equations (\ref{transf1}) and (\ref{transf2}) through Killing equations $\delta\psi_m=0$ and $\delta\lambda_A=0$ preserving some supersymmetry \cite{gibbons-lambert}, leading respectively to
\be
A'=\mp 2\kappa^{D-2} W, \qquad {\phi^A}'=\pm 2 G^{AB}W_{3B},
\ee
then we can write the following first-order differential equations  (for upper signs) \cite{Bazeia:2007vx, DeWolfe}
\bea
\label{5d-metric-A}
A'&=&-\frac{W}{3M_*^3},\\
\label{5d-phi}
\phi'&=&W_\phi, \\
\label{5d-chi}
\chi'&=&W_\chi,
\eea
where we have properly rescaled the superpotential to absorb $2(D-2)$ factors and considered $D=5$. The scalar component fields have been chosen as $\phi_1=\phi/2$ and $\phi_2=\chi/2$. Finally, the subscripts $\phi, \chi$ stand for derivatives with respect to these fields.

The most general form of the superpotential that generates scalar potential with a $Z_{2}\times Z_{2}$-symmetry  \cite{JHEP} is given by
\bea\label{4}
	W=\lambda\left(\frac{\phi^3}{3}-a^2\phi\right)+\mu\phi\chi^{2}. 
\eea

} The Bogomol'nyi-Prasad-Sommerfield (BPS) solutions of the first-order differential equations (\ref{5d-phi})-(\ref{5d-chi}) are of type I
\bea\label{6}
	&\phi=-a\tanh(\lambda ax_5)& \\
	&\chi=0 \nonumber&
\eea  
and type II
\bea\label{7}
	&\phi=-a\tanh(2\mu ax_5)& \\
	&\chi=\pm a\sqrt{\frac{\lambda}{\mu}-2}{\;\rm sech}{(2\mu ax_5)}& \nonumber
\eea 
where $\pm a$ are the minima of the potential. The type I solution is not interesting for our proposal since it produces reflectionless domain walls --- see below.

{ {We shall now consider the limit $W/3M_*^3\ll1$ into (\ref{5d-metric-A}), in which the 3D domain walls are solutions embedded in 5D Minkowski space (\ref{5d-metric}) with $A=const.$ Alternatively, for computing the induced inter brane force shown below, we shall assume, without loss of generality, the limit where the gravitational field in the bulk is considered a weak background field, such that the branes are essentially living in a Minkowski space  \footnote{The four-dimensional gravity can be induced on the brane via  DGP mechanism  \cite{Dvali:2000hr,Dvali:2000xg}, i.e., through quantum loops of matter fields localized on the brane.}.

Thus, in such a flat limit the analysis to find supersymmetric domain walls from the Lagrangian (\ref{1}) reduces to work with an effective Lagrangian as in the following \cite{RIBEIRO, D61}} 
\bea\label{2}
	{\cal L}=\frac{1}{2}\partial_{M}\phi\partial^{M}\phi+\frac{1}{2}\partial_{M}\chi\partial^{M}\chi-V(\phi,\chi), 	
\eea 
whose equations of motion are
\bea\label{EOM}
\square\phi+\frac{\partial V}{\partial\phi}=0, \nonumber\\
\square\chi+\frac{\partial V}{\partial\chi}=0,
\een 
{with the scalar potential now given by}
\bea\label{3}
	V(\phi,\chi)=\frac{1}{2}\left(\frac{\partial W}{\partial\phi}\right) ^{2}+\frac{1}{2}\left(\frac{\partial W}{\partial\chi}\right)^2.
\eea  
{ 
Substituting Eq.(\ref{4}) into Eq.(\ref{3}), we can write the scalar potential as
\bea\label{5}
	V(\phi,\chi)=\frac{1}{2}\lambda^{2}(\phi^{2}-a^2)^{2}+(2\mu^{2}+\lambda\mu)\phi^{2}\chi^{2}-\lambda a^{2}\mu\chi^{2}+\frac{1}{2}\mu^{2}\chi^4. 
\eea

\subsection{The domain wall reflection coefficient} 

}

Now performing small perturbations around a particular solution, say, $\bar{\phi}$ and $\bar{\chi}$, that is \bea\label{8}
	&\chi=\bar{\chi}+\zeta &\\
	&\phi=\bar{\phi},& \nonumber
\eea
in the equations of motion (\ref{EOM}), then we obtain a linear equation for the fluctuations 
\bea\label{zeta}
\partial_\mu \partial^\mu \zeta + \bar{V}_{\chi\chi}\zeta=0.
\een
For type II solution above \cite{Brito,JHEP} we have
\bea
\bar{V}_{\chi\chi}(x_5)=m_\chi^2-m_\chi^2 \left(4-\frac{\lambda}{\mu}\right){\rm\, sech}^2( 2\mu a x_5)
\een
with $m_\chi^2=4\mu^2a^2$ being the mass squared of the elementary excitations of the scalar field $\chi$. The {\it Ansatz} for the perturbation around a three-dimensional domain wall can be chosen as
\bea
\zeta=\zeta(x_5)e^{-i(\omega t-k_x x-k_yy-k_zz)}.
\een
Substituting this into the equation (\ref{zeta}) we find the Schroedinger-like equation
\bea\label{sch-zeta}
-\frac{d^2\zeta}{dx_5^2}+U_{II}(x_5)\zeta=k_5^2\zeta
\een
with $-k_5^2=-\omega^2+k_x^2+k_y^2+k_z^2+m_\chi^2$. Here $k_5$ is the fifth-component bulk particles momentum and  
\bea\label{9}
	{U_{II}}(x_5)=-m^{2}_{\chi}\left(4-\frac{\lambda}{\mu}\right)\;{\rm sech}^{2}{(2\mu ax_5)}
\eea
is the Schroedinger-like potential. The reflection coefficient \cite{Vilenkin} for the barrier potential described in Eq.~(\ref{9}) is
\bea\label{10}
	R=\frac{\cos^{2}(\frac{\pi}{2}\sqrt{17-\frac{4\lambda}{\mu}})}{\sinh^{2}(\frac{\pi k_{5}}{2\mu a})+\cos^{2}(\frac{\pi}{2}\sqrt{17-\frac{4\lambda}{\mu}})}
\eea
Notice that for $\lambda=2\mu$, that reduces the solution type II (\ref{7}) to type I (\ref{6}), the reflection coefficient becomes zero.

\begin{figure}[h]
\centering{\includegraphics[scale=0.4]{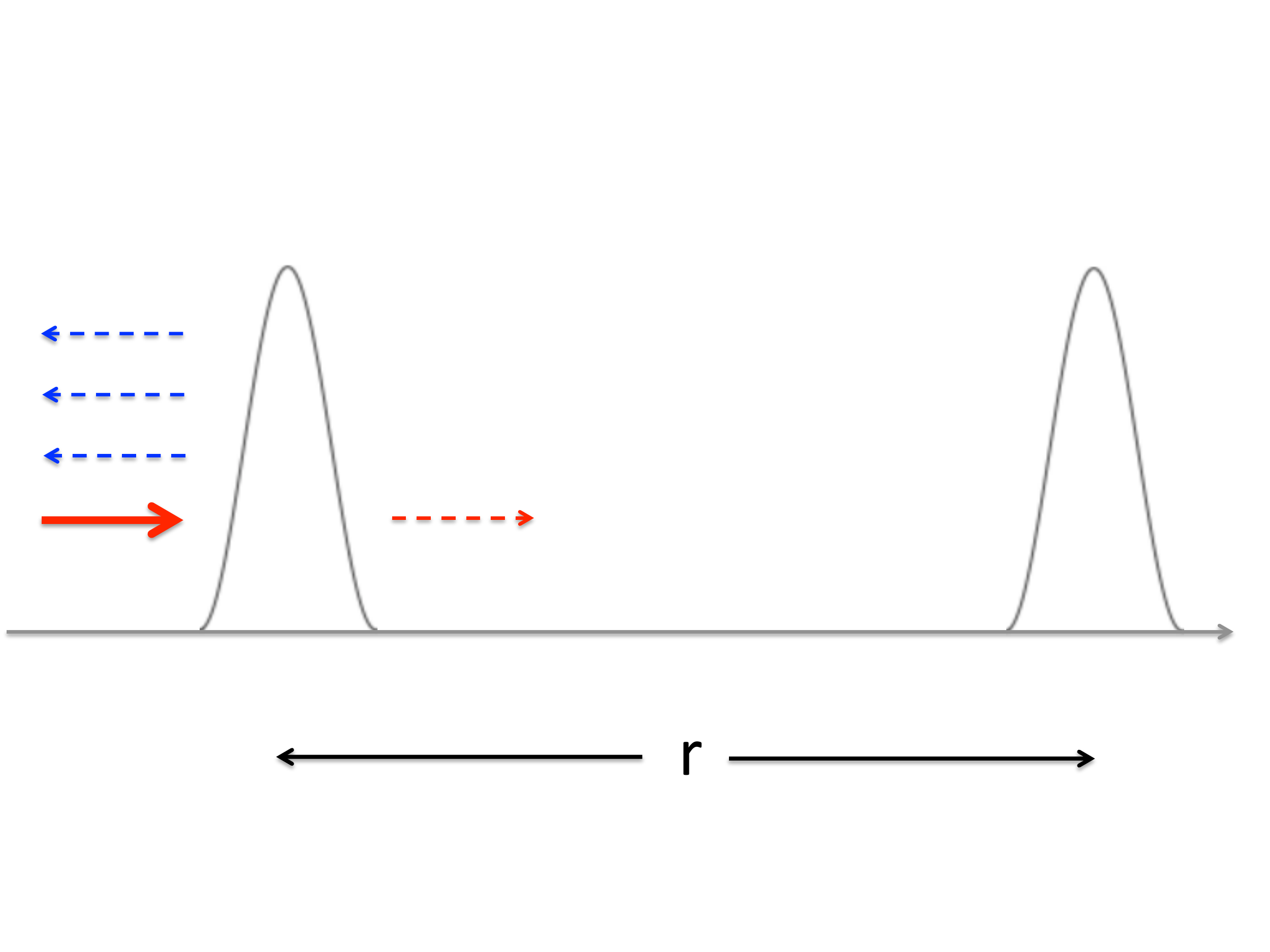}}
\centering{\includegraphics[scale=0.4]{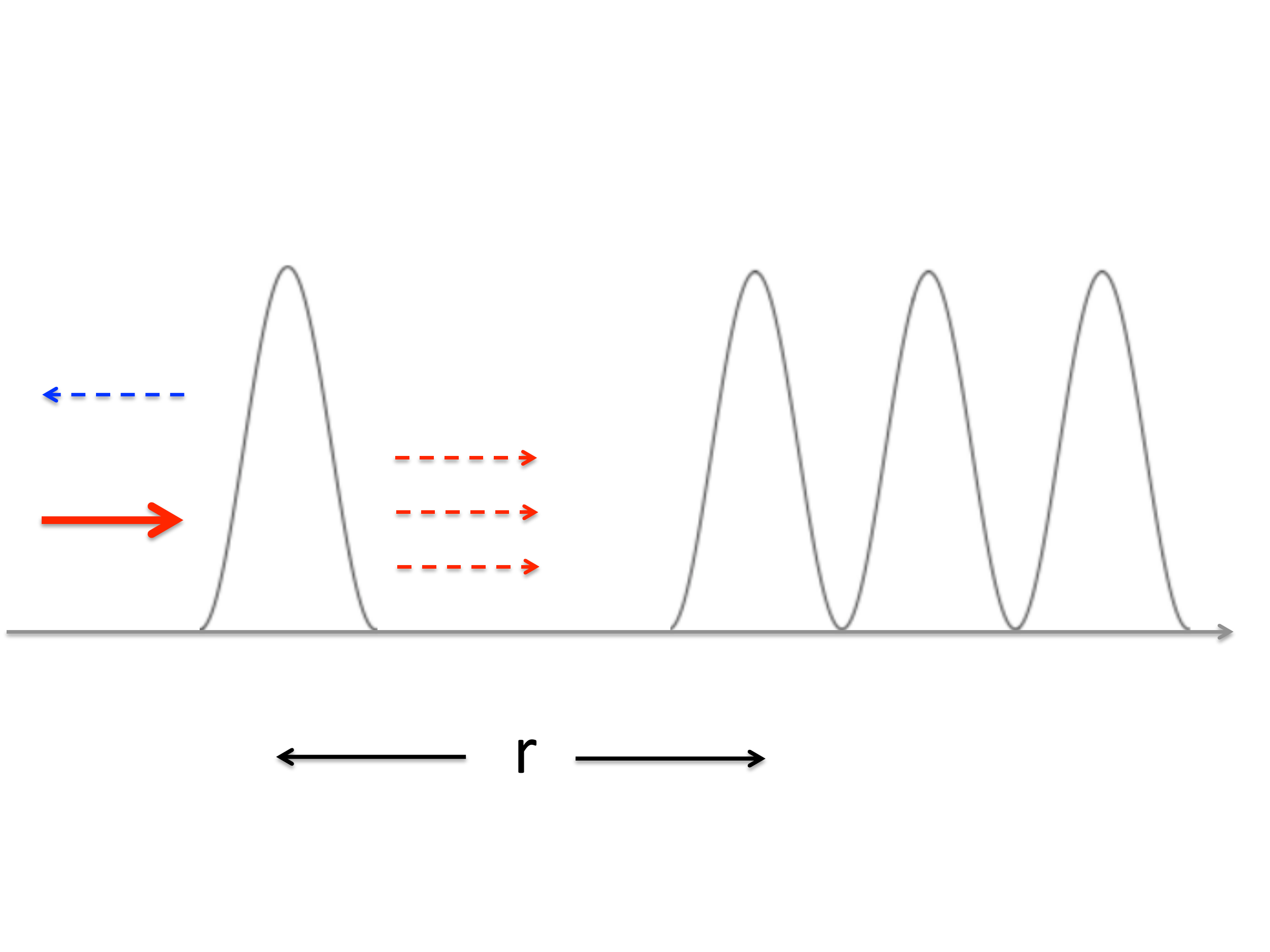}}
\caption{Reflecting particles in multiple barrier potentials. (Top panel) The left barrier is sufficiently distant from the the right barrier and does not experiment the {\it resonant tunneling effect} \cite{Tye:2009rb,mzbacher} which would diminish the reflection rate. (Bottom panel) The left barrier is sufficiently near, at a distance $r$, to an array of barrier on the right. Now the resonant tunneling effect takes place in between favoring the transmission rate in contrast to the reflection rate. Then it is expected to exist an attractive force between two branes whose magnitude increases (decreases) as the inter distance $r$ increases (decreases).}\label{figs-N-branes}
\end{figure}

\eject

\section{The induced inflaton potential}\label{inflaton}

So by considering the transverse force along the inter distance $r$ in the fifth coordinate due to elastic collisions we have \cite{Vilenkin}
\bea\label{11}
	F_{r}=M_{wall}\ddot{r}(t)\simeq KR=-\frac{\partial U}{\partial r},
\eea
where $K$ depends on the density of colliding bulk particles and their incoming momenta $k_5$, that we shall assume to be time independent. This is expected because of the conservation of momentum of both colliding particles and the brane. We can compare this to a similar computation such as the recoil of nuclei in alpha decay due to a specific barrier potential. The essential difference concerns to the fact that, while in the latter case the recoil is usually disregarded since one assumes heavy nuclei, in the former case we shall assume that the conservation of momentum involves motion of both particles and branes.  We can obtain the potential $U(r)$ from equation (\ref{11}) as long as we are able to find the reflection coefficient as a function of the inter distance $r$, i.e., $R(r)$. This is indeed the case if we take in consideration a second parallel brane put near the first brane --- see Fig.~\ref{figs-N-branes} (top). As such the reflection coefficient changes because of the {\it resonant tunneling effect} \cite{Tye:2009rb,mzbacher}, Fig.~\ref{figs-N-branes} (bottom),  where the transmission coefficient is given by \cite{mzbacher}
\bea
T=\frac{4}{\left(4\theta^2+\frac{1}{4\theta^2}\right)\cos^2{L}+4\sin^2{L}},
\eea
where \footnote{Here we have adapted dimensions in the Schroedinger-like equation (\ref{sch-zeta}) by dividing each term by a scale of mass $2m$. This implies on ${\cal V}(x_5)=U_{II}(x_5)/2m$ and $E=k_5^2/2m$. This also happens with the resonance energy $E_0=k_0^2/2m$ --- see below. For the sake of convenience, however, we also have made use of the convention $\hbar=1$ and $2m=1$.}
\bea
\theta=\exp{\left( \int_a^b \kappa(x_5)dx_5\right)}, \qquad \kappa(x_5)=[{\cal V}(x_5)-E]^{1/2}, \qquad E<{\cal V}(x_5),
\een
gives the hight and thickness of the barrier Fig.~\ref{figs-N-branes} in terms of the energy 
and 
\bea
L=\int_{-r}^r{k(x_5)dx_5}, \qquad k(x_5)=[E-{\cal V}(x_5)]^{1/2}, \qquad E>{\cal V}(x_5).
\een
In the limit of pronounced resonances ($\theta\gg1$) one may assume around the resonances the approximation $\cos{L}\approx\mp (\partial L/\partial E)_{E=E_0}(E-E_0)$ and 
$\sin{L}\approx1$. Then we can still write the transmission coefficient as
\bea
T=\frac{(\Gamma/2)^2}{(E-E_0)^2+(\Gamma/2)^2}
\eea
where by definition $\Gamma=(\theta^2  (\partial L/\partial E)_{E=E_0})^{-1}$.
Now by assuming the Schroedinger-like potential ${\cal V}(x_5)$ in between the barriers is sufficiently small (which is true for branes sufficiently far from each other) we can find the reflection coefficient $R=1-T$, given in the form
\bea\label{R2}
R=\frac{(E-E_0)^2}{(E-E_0)^2+(\Gamma/2)^2}\to R(r)=\frac{1}{1+\frac{r_0^2}{r^2}}
\eea
Since $E-E_0\ll1$, in the last step above we have recast the formula in terms of the fixed distance scale $r_0=k_0/[2\theta^2(E-E_0)]$ and also used $ (\partial L/\partial E)_{E=E_0}=r/k_0$.
Now substituting (\ref{R2}) into equation (\ref{11}), we can integrate $R(r)$ to find the potential that acts in between the parallel branes
\bea\label{13}
	U(r)=Kr_0\arctan\left(\frac{r}{r_0}\right)-Kr
\eea
However, due to the existence of a linear potential contribution as a consequence of constant electric and gravitational fields between the branes \cite{Dvali}, the total potential governing the motion of such branes is indeed
\bea\label{14}
	U_{eff}(r)=U(r)+{\cal E}_{0}r
\eea
Thus for ${\cal E}_{0}=K$, we can write the effective potential
\bea\label{15}
	U_{eff}(r)=Kr_0\arctan\left(\frac{r}{r_0}\right)
\eea

In the following we shall show how the electric field enters in the present scenario. This can be well justified by introducing gauge fields contribution into the Lagrangian (\ref{2}) as follows \cite{Brito:2012gp,Bazeia:2016pra}:
\bea
{\cal L}=-\frac14F_{MN}F^{MN}+J_M A^M+\frac{1}{2}\partial_{M}\phi\partial^{M}\phi+\frac{1}{2}\partial_{M}\chi^*\partial^{M}\chi-V(\phi,|\chi|). 
\eea
Here we have promoted the scalar field $\chi$ to be a complex field $\chi(x_5,x_\mu)=\chi(x_5)\exp(i\theta_\mu x^\mu)$ in order to describe charged domain walls with the current $J_M=iq(\chi\partial_M\chi^*-\chi^*\partial_M\chi)=(J_\mu,0)$, where $J_\mu=-q\theta_\mu\chi(x_5)^2$, $q$ is the electric charge and $\mu=0,1,2,3$ labels the brane world-volume coordinates. For static gauge fields with translational symmetry along the brane embedded in 5D Minkowski space, the Gauss law simply reduces to
\bea
\frac{d^2U}{dx_5^2}=\rho(x_5),
\een
where $A^0=U$, $\vec{A}=0$, and $\rho=J_0$. Now by using equation (\ref{7}) for the solution $\chi(x_5)$ we find the charge density on two parallel domain walls located at $\pm\, r/2$ as 
\bea
\rho(x_5)=\frac{\sigma}{2\Delta}{\rm sech}^2{\left(\, \frac{x_5\pm r/2}{\Delta}\right)},
\een
where $\Delta\sim 1/ (2\mu a)$ is the domain wall thickness and $\sigma=q/\Delta^2$. We also have eliminated the parameter $\theta_0$ in terms of the parameters $\lambda, \mu, a$ that appear in the amplitude of the solution for $\chi$ (\ref{7}). It is not difficult to show that the electric field between two parallel domain walls with opposite charge is 
\bea
\frac{dU}{dx_5}=\int{dx_5\Big(\rho(x_5+r/2)-\rho(x_5-r/2)\Big)}=\frac{\sigma}{2}\left(\tanh{\frac{x_5+ r/2}{\Delta}}-\tanh{\frac{x_5- r/2}{\Delta}}\right),
\eea
where it clearly approaches to a constant electric field for large distance $r$ and becomes zero as they overlap, i.e., at $r=0$, as expected. The potential is obtained by integrating the electric field in the interval $(-r,r)$ to find 
\bea
U=\frac{\sigma}{2}\Delta\ln\left.\left(\cosh\frac{x_5+r/2}{\Delta}{\rm\, sech}{\, \frac{x_5- r/2}{\Delta}}\right)\right|_{-r/2}^{r/2}\to U={\cal E}_0\,r,
\een
where in the last step we have taken the large distance limite $r\gg \Delta$ and defined ${\cal E}_0=\sigma$. This is precisely the linear term added to the `effective potential' given in (\ref{14}).

Now considering the total energy of the brane per unit 
volume, we find
\bea\label{16}
	\frac{E}{V_{(3)}}=\frac{1}{2}\frac{M_{wall}}{V_{(3)}}\dot{r}^{2}+\frac{U_{eff}}{V_{(3)}},
\eea
where $\rho_{wall}={E}/{V_{(3)}}$ is the energy density of the 3-brane, $T_{wall}={M_{wall}}/{V_{(3)}}$ is the tension and $V(r)={U_{eff}}/{V_{(3)}}$ is the potential density in four dimensions.
Then 
\bea\label{17}
	\rho_{wall}=\frac{1}{2}T_{wall}\dot{r}^{2}+V(r)
\eea
and admitting that $\sqrt{T_{wall}}\,r(t)\longleftrightarrow\phi(t)$ we find the total energy density 
\bea\label{18}
	\rho_{wall}=\frac{1}{2}\dot{\phi}^{2}+V(\phi),
\eea
where 
$\phi=\phi(t)$ is denominated \textit{inflaton} whose associated potential is
\bea\label{29}
	V(\phi)=K\beta\arctan\left(\frac{\phi}{\beta}\right),
\eea
with $\beta=\sqrt{T_{wall}}\,r_0$ describes the scale of energy. This induced potential will drive the inflationary scenario discussed in the next section.

\section{Inflationary Cosmology on the Brane}\label{sec02}

The central idea of scalar field inflation models is to consider that the energy of the early Universe has been dominated by the potential energy of  scalar fields. The parameters that characterize the slow-roll 
\bea\label{25}
	\epsilon=\frac{M^{2}_{Pl}}{16\pi}\left(\frac{V'(\phi)}{V(\phi)}\right)^{2}
\eea
and
\bea\label{26}
	\eta=\frac{M^{2}_{Pl}}{8\pi}\left(\frac{V''(\phi)}{V(\phi)}\right)
\eea
are valid as long as both are small $\left(\epsilon\ll1, \eta\ll1\right)$ \cite{Weinberg}.
The spectral index $n_{s}$ and the tensor-to-scalar ratio $r$ are given in terms of these parameters as follows
\bea\label{27}
	n_{s}=1+2\eta-6\epsilon
\eea
and
\bea\label{28}
	r=16\epsilon
\eea 
The tensor-to-scalar ratio $r$ measures how much the tensor perturbations change with the scale and according to more recent data has the upper bound $r<0.07$ \cite{Planck, Planckk} and $r<0.02$ \cite{Plan}.

The main results can be summarized as follows. To work with Eqs.(\ref{27}) and (\ref{28}) we should analyze the behavior of the scalar field $\phi$ and cosmological parameters in the slow-roll regime, that is, $\phi\gg\beta$. This precisely happens in the {\it flat region} of the potential (\ref{29}). 
The inflationary phase is maintained as long as $\epsilon(\phi)\ll1$ and ends as  $\epsilon(\phi)\sim1$. 
The number of $e$-folds for the slow-roll approximation can be obtained as a function of the scalar field and is found by using 
\bea\label{38}
	{ N}=\frac{8\pi}{M^{2}_{Pl}}\int_{\phi_{end}}^{\phi}{\frac{V(\phi)}{V_{\phi}(\phi)}d\phi}.
\eea
{From the equation (\ref{25})-(\ref{26}) and potential (\ref{29}) at the limit  $\phi\gg\beta$ we find 
\be\label{N-end}
N\simeq \frac{8\pi\beta^2}{M_{Pl}^2}\left(\frac{\phi_N^3}{3\beta^3}-\frac{\phi_{end}^3}{3\beta^3}\right).
\ee
For $\epsilon$ or $|\eta|\sim1$ the inflation ends. At the approximation considered above, $|\eta|\sim1$ gives 
\be
\left(\frac{\beta^2}{\phi^2_{end}+\beta^2} \right)^2\sim \frac{\beta^2}{M_{Pl}^2}, 
\ee
and then we find $\phi_{end}\sim (M_{Pl}\beta)^{1/2}$. Now substituting this into (\ref{N-end}) we achieve the following relationship
\be\label{N-phi-end}
\phi_N\sim\phi_{end}\left[1+\frac{3}{8\pi}\left(\frac{M_{Pl}}{\beta}\right)^{1/2}N\right]^{1/3}.
\ee
Further consideration about this result will be considered below.
}

{The parameter $K$ of the potential (\ref{29}) can be determined at the pivot scale $k_*$ (scale at which CMB crosses the Hubble horizon during inflation) as follows
\be
P_R=\left.\frac{V(\phi)}{24\pi^2\epsilon M_P^4}\right|_{k=k_*}
\ee
that for $k_*=0.05$ Mpc$^{-1}$, $P_R(k_*)=A_s$ determined up to Planck 2018 normalization $A_s=2.0933 \times 10^{-9}$ and using the equations for the potential (\ref{29}) and $\epsilon$ (\ref{25}) one can solve for $K$ to obtain
\be\label{amp-V}
K= \frac{12M_{Pl}^2\pi^2P_R(k_*)M_P^4}{\beta^3\arctan^3{\left(\frac{\phi_*}{\beta}\right)}\left(1+\frac{\phi_*^2}{\beta^2}\right)^2}.
\ee
We can show that $\phi_N\sim 4\phi_{end}$ for $50\leq N\leq 70$ from (\ref{N-phi-end}) by using the fact that $\beta=10^{-2}M_{Pl}$ according to the Planck data and the constraining of the parameters as shown in table \ref{Tabela 1}. 
Since at pivot scale $\phi_*\sim\phi_N$, then we can assume ${\phi_N}\sim\phi_*\sim\phi_{end}$. Thus, we can also approximate the equation (\ref{amp-V}) by using $\phi_*\gg\beta$ as follows
\be\label{amp-V2}
K\beta= \frac{12M_{Pl}^2\pi^2P_R(k_*)M_P^4}{\beta^2\left(\frac{\phi_*^2}{\beta^2}\right)^2}\sim\frac12P_R(k_*)M_P^4,
\ee
where in last step we have used $\phi_*^2\sim16\,\phi_{end}^2=16(M_{Pl}\beta)$. Recall that $M_P=M_{Pl}/(8\pi)^{1/2}$ is the reduced Planck mass $\sim 10^{18}$ GeV. We may now estimate the reheating temperature by assuming that the initial potential energy density $V\sim K\beta$ related to the inflaton field is of the same order of the converted energy to radiation and ultra-relativistic particles with energy density $(\pi^2/30)g_*T^4$ as follows
\be
T_{rh}\sim\left(\frac{30}{\pi^2}\frac{K\beta}{g_*}\right)^{1/4}\sim 10^{15}\, {\rm GeV},
\ee
where $g_*\sim106$ is the number of effectively massless species for temperature above $300$ GeV according to the standard model \cite{kolb-turner}. 

} 

\begin{table}[h]
\centering
\caption{68\% confidence limits for the cosmological parameters using the TT+lowE Planck (2018).}
\label{Tabela 1}
\vspace{0.4cm}
\begin{tabular}{cccc}
\hline
Parameter & $\Lambda$ CDM & brane-infla. ${ N} = 50$ & brane-infla. ${N} = 60$ \\
\hline
$n_{s}$ & 0.9626 $\pm$ 0.0057 & 0.9600 $\pm$ 0.0027 & 0.9774 $\pm$ 0.0015 \\
$r_{0.002}$ & $<0.102$ & 0.0141 $\pm$ 0.0012 & 0.0068 $\pm$ 0.0013 \\
$\beta$ & - & $0.032\pm0.001$ & $0.032\pm0.001$ \\
\hline
\end{tabular}
\end{table}

The Tab. \ref{Tabela 1} shows the values of the cosmological parameters obtained from the analysis of the model
for $\beta$ given in $M_{Pl}$. 
The first column shows constraints on the $\Lambda$CDM reference model (TT + lowE 68\% confidence) verified from the observations of the Planck and Collaborations 2018 \cite{Plan}, the second and third columns respectively show the values of $n_{s}$ and $r$ for ${ N} = 50$ and ${N} = 60$ obtained in our model.

\section{Conclusions}\label{sec04}

In this work, we use field theory to construct a cosmological model obtained through the brane inflation scenario, constructed by assuming the possibility of the Universe is living on a thick brane that moves toward a stack of branes due to collision of particles that live in a five-dimensional bulk. From this model, we analyze some aspects of the theory of inflation governed by a scalar field that assumes the role of inflaton.  
We find some cosmological parameters to compare with the last data of Planck and Collaborations \cite{Plan}. All the values of parameters we find are in agreement with the currently expected data for the inflationary era.

\acknowledgments

We would like to thank CNPq, CAPES and PRONEX/CNPq/FAPESQ-PB (Grant no. 165/2018), for partial financial support. FAB acknowledges support from CNPq (Grant no. 312104/2018-9). The authors also thank A.R. Queiroz and S. S. Costa for discussions.

\end{document}